\documentclass{aastex631}
\usepackage{amsmath}
\usepackage{array}
\usepackage{amsmath}
\usepackage{hyperref}

\begin{document}

\title{Cluster Catalog Validation with Multiwavelength Data}

\author{E.\ K.\ Steelberg}
\affiliation{Department of Astronomy and Astrophysics, University of California, Santa Cruz, CA 95064, USA}

\author{T.\ E.\ Jeltema}
\affiliation{Department of Astronomy and Astrophysics, University of California, Santa Cruz, CA 95064, USA}
\affiliation{Santa Cruz Institute for Particle Physics, Santa Cruz, CA 95064, USA} 

\author{J.\ H.\ O'Donnell}
\affiliation{Department of Astronomy and Astrophysics, University of California, Santa Cruz, CA 95064, USA}
\affiliation{Santa Cruz Institute for Particle Physics, Santa Cruz, CA 95064, USA} 

\author{R.\ Solomon}
\affiliation{LAPP, Université Savoie Mont Blanc, CNRS/IN2P3, Annecy; France}

\author{The LSST Dark Energy Science Collaboration}

\begin{abstract}
\setlength{\parindent}{0pt}
\small

The Legacy Survey of Space and Time (LSST) will provide a ground-breaking data set for cosmology, but to achieve the precision needed, the data, data reduction, and algorithms measuring the cosmological data vectors must be thoroughly validated and calibrated. In this note, we focus on clusters of galaxies and present a set of validation tests for optical cluster finding algorithms through comparison to X-ray and Sunyaev-Zel'dovich effect cluster catalogs. As an example, we apply our pipeline to compare the performance of the redMaPPer (red-sequence Matched filter Probabilistic Percolation) and WaZP (Wavelet Z Photometric) cluster finding algorithms on Dark Energy Survey (DES) data.

\end{abstract}

\section{Introduction}
\small
LSST will discover hundreds of thousands of clusters giving the potential for precise constraints on cosmology if cluster selection and mass calibration are well understood.  In addition, the data processing and photometry must be validated, since clusters are special environments with high galaxy densities and massive galaxies with extended stellar envelopes.  Comparison of optically-selected clusters to multiwavelength data has proven effective at diagnosing issues in early catalogs \citep[e.g.][]{DESY6Gold} and at calibrating aspects of cluster selection \cite[e.g.][]{2024}.
 
In this paper, we introduce a validation pipeline which cross-correlates cluster catalogs with multiwavelength data and implements a series of performance tests. While developed for LSST-DESC, this pipeline can be broadly applicable. 
\section{Methodology}

Our pipeline performs a series of validation tests of a user input cluster catalog, including completeness, centering, and observable scatter compared to X-ray and SZ data, and outputs summary plots, statistics, and comparisons to literature values.  It is available on GitHub as an easy to run python notebook\footnote{https://github.com/LSSTDESC/Initial-Cluster-Catalog-Validation} along with a detailed description of all validation tests and summary figures that the pipeline provides.  

\subsection{Catalogs and Matching}

As an example, we apply our pipeline to the publicly available DES Y3 redMaPPer \citep{Rykoff_2014} and the DES Y1 WaZP \citep{Aguena_2021} cluster catalogs. We set richness cuts of $\lambda > 20$ and $N_{gals} > 25$ for redMaPPer and WaZP respectively. We also limit the redshift range to $0.2 < z < 0.65$ for all proceeding analysis, matching the choices used for DES cluster cosmology \citep{DESY3cosmo}. 

The input cluster catalog is compared against two multiwavelength catalogs. 
 The first is the 2500 $\text{deg}^2$ microwave catalog of clusters detected via the thermal Sunyaev–Zeldovich effect from the South Pole Telescope (SPT) \citep{spt2500}, which is used to test completeness and observable scatter. We cut the SZE-SNR above $\xi >5$ as high significance clusters should be found by each selection algorithm. As such, the completeness of the optical catalog is not determined overall; however, low recovery may indicate gaps within the catalog.  The second is a curated X-ray catalog based on the RASS-MCMF cluster catalog \citep{MCMF}. This catalog is used to test centering performance and scatter but not completeness, and thus we select only clusters with well-determined X-ray centers and consistently determined X-ray temperatures. Specifically, we use only RASS-MCMF clusters with archival Chandra observations and process these with the MATCha pipeline \citep{matcha} to determine X-ray temperatures and X-ray peak centers. We further limit this catalog to clusters with SNR$>25$, $z>0.1$, Dec $<20$ deg, and we visually remove disturbed clusters and those near chip edges. Since for our test of mis-centering we match only to well-centered X-ray clusters, we do not expect the mis-centering parameters to reflect the optical catalogs overall. However, significant mis-centering may indicate problems in the optical selection.

We perform cross-matching of the input and SZ/X-ray catalogs using the ClEvaR (Cluster Evaluation Resources) Python library developed by \cite{CLEvaR}.  Optionally, survey masks can be applied to each of the catalogs.  For our examples in this paper, we take the footprint from the Y1 cosmology redMaPPer cluster catalog and select only the clusters in SPT that are found within this mask. Each cluster catalog is loaded into the cross-matching function which inputs RA, Dec, redshift, and richness. The matching parameters are set according to users specification by defining radial and redshift offsets. We then specify a preference of which cluster is picked in the case where a cluster matches to multiple objects in the multi-wavelength data. In our analysis, we set a preference on the more massive cluster based on the input observable (e.g.~richness). 
Matched clusters are saved and outputted into a separate catalog, where duplicate matches found initially are also recorded. Here we use a 2 Mpc separation and a redshift offset of 0.05.

\subsection{Validation Tests}

The validation pipeline runs a series of three tests on matched clusters, completeness compared to SPT SZ, centering performance compared to X-ray, and scatter of the input observable (e.g.~richness) compared to X-ray temperature and SZ detection significance $\xi$. It then outputs a series of summary plots and statistics.

First, the code plots the recovery fraction of SZ clusters as a function of SZ significance $\xi$.   

To test the centering performance, we calculate the offsets between the optical and X-ray centers of the matched optical to X-ray clusters.  The distribution of offsets is fitted to a two-component gamma distribution following \cite{2019} where $\rho$ is the fraction of well-centered clusters and $\sigma$ and $\tau$ are the scales of the well-centered and mis-centered distributions, respectively.

We then run a Markov-Chain Monte Carlo fit on the cross-matched X-ray to optical position offsets. The code outputs the best-fit $\rho, \sigma$, and $\tau$. 
Functions for plotting summary figures including corner plots of parameter constraint distributions and a histogram of X-ray peak to optical position offsets with the best-fit model overlaid are provided. The histogram plot also includes a comparison to the well-centered and mis-centered models found by \cite{2024} for joint XMM and Chandra X-ray follow up of DES Y3 redMaPPer clusters. 

The final validation tests fit the scaling relations between cluster-mass observables. We investigate the scaling and intrinsic scatter of X-ray temperature and SZ significance with respect to redMaPPer/WaZP richness. Our scaling relations have the form $\ln (y)  = \alpha \ln(x) + \beta$ where $\alpha , \beta,$ and $\sigma$ are the slope, intercept, and scatter found by implementing the Bayesian regression model in \cite{2007}. To fit these scaling relations, we run code from the \href{https://github.com/sweverett/CluStR}{CluStR} python package \citep{2024} and fit for both ($E(z)^{-\frac{2}{3}}k_{B}T_{X}(r_{2500}) - \text{Richness}$) and ($\text{Richness} - \xi$) .

\section{Results}

Table 1 shows example results using the validation code for DES Y3 redMaPPer and DES Y1 WaZP.

\begin{table}[h]
    \centering
    \begin{tabular}{ccccccccc}
         Sample & $\rho$ & $\sigma$ & $\tau$ & $\sigma$ & $\sigma$ & $N_{spt}$ & $N_{xray}$ & \parbox[c]{2cm}{Completeness} \\ 
         & & & & ($T_X-\lambda/N_{gals}$) & ($\lambda/N_{gals}-\xi$) & & &\small{$\xi > 5$}\\
         \hline
         
         redMaPPer & $0.94 \pm 0.07$ & $0.067 \pm 0.013$ & $0.32 \pm 0.21$ & $0.27 \pm 0.03$ & $0.33 \pm 0.02$ & 134 & 63 & $ 88.7\%$ \\ 
         
         WaZP & $0.92 \pm 0.07$ & $0.050 \pm 0.013$ & $0.34 \pm 0.20$ & $0.29 \pm 0.05$ & $0.46 \pm 0.03$ & 138 & 29 & $91.4 \%$\\ 
         
         Kelly et al. (2024) & $0.87 \pm 0.04$ & $0.053 \pm 0.006$ & $0.23 \pm 0.05$ & $0.22\pm 0.01$ & N/A & N/A & 243 & N/A\\ 

    \end{tabular}
    
    \caption{Best fit values and 1$\sigma$ uncertainties for the centering parameters, intrinsic scatter of observable scaling relations indicated, the number of matched clusters, and the completeness compared to SPT SZ for $z \in (0.2, 0.65)$ of DES Y3 redMaPPer and DES Y1 WaZP, respectively}
    \label{tab:results}
\end{table}

Overall, out of 151 SPT selected clusters in our chosen redshift range and footprint, we find 134 matches from redMaPPer and 138 matches from WaZP. Matching analysis indicates that both redMaPPer and WaZP show excellent completeness with respect to the SPT sample of clusters, with 100\% recovery of SPT clusters above $\xi \sim10$.  It is unlikely that chance matches contributed to the high completeness, as the distribution of SPT clusters is sparse compared to the DES footprint.

Well-centered fractions and widths of centered and miscentered distributions are shown in Table \ref{tab:results}.
Both redMaPPer and WaZP demonstrate similar centering performance. These results are consistent with what we expect for our well-centered X-ray sample and with previous analysis of redMaPPer performance on Chandra and XMM clusters from \cite{2024}. 
  
The origins of miscentering in cluster selection can be attributed to multiple factors. Masking due to the presence of bright stars or other features in the data can lead to miscentering as well as confusion or ambiguity in merging clusters with substructures that host bright galaxies.  Additionally, central galaxies with AGN or star formation that lead to a galaxy color that is bluer than expected can lead redMaPPer to miss the central galaxy.
Overall, however, our results indicate that both cluster finders identify the correct center in $\sim 93$\% of clusters with unambiguous X-ray centers.

The results for the scatter in the ($E(z)^{-\frac{2}{3}}k_{B}T_X(r_{2500})-\text{Richness}$)  and ($\text{Richness} -\xi$) relations are presented in Table \ref{tab:results}. We find both redMaPPer and WaZP show agreement in the scatter of the ($E(z)^{-\frac{2}{3}}k_{B}T_X(r_{2500})-\text{Richness}$) relation for the limited sample of matched clusters from ClEvaR. In both cases our scatter was marginally higher then the sample used in \cite{2024}, but not significantly so. In our ($\text{Richness} - \xi$) relation we note that the scatter for WaZP was larger than for redMaPPer. 

\section{Conclusion}
We have developed a set of validation tests for optically-selected cluster catalogs using multiwavelength data, and demonstrate these on public DES cluster catalogs. Our validation tests demonstrate good centering performance of both redMaPPer and WaZP. In both cases 8\% or less of the clusters were found to be miscentered.
The intrinsic scatter of cluster-mass observables in both algorithims have low scatter compared to the X-ray catalog. For comparison with SZ, WaZP had a higher scatter than redMaPPer.

While here we use as examples final, published cluster catalogs, we note that earlier versions of DES cluster catalogs have had features or bugs that would be revealed by these tests. Implementation of further validation tools to test the effects of survey masking are planned for the future. Such developments will increase our ability to quickly diagnose issues that may arise with cluster finding in LSST.

\section*{Acknowledgments}
DESC acknowledges ongoing support from the IN2P3, the STFC, and the DOE and LSST DA. DESC uses resources of the IN2P3, NERSC, and DiRAC HPC, and GridPP computing facilities. This work was performed in part under DOE contract DE-AC02-765F00515.

\bibliographystyle{aasjournal}
\bibliography{bibfile}

\end{document}